\documentclass[aps,pra,reprint,groupedaddress,final,showpacs]{revtex4-1}

\usepackage[utf8]{inputenc}
\usepackage{graphicx}
\usepackage{color}
\usepackage{amsmath}
\usepackage{times}

\def\pra{Phys. Rev. A~}

\def\prl{Phys. Rev. Lett.~}

\def\reff#1{(\ref{#1})}

\def\diff{\mathrm{d}}
\def\imagi{i}

\def\beq{\begin{equation}}
\def\eeq{\end{equation}}

\def\Tp{{T_{\mathrm{p}}}}

\def\ket#1{\vert #1\rangle}
\def\bra#1{\langle#1\vert}

\def\eulere{\,e}

\def\Wcmcm{W/cm$^2$}

\begin{document}

\title{Dissociative ionization of H$_2^+$: Few-cycle effect in the joint electron-ion energy spectrum}
%\title{Joint electron-ion energy spectrum for dissociative ionization of H$_2^+$ reveals few-cycle effect}
%\title{Few-cycle effect in dissociative ionization of H$_2^+$}

\author{V.\ Mosert}
\author{D.\ Bauer}
%\email[]{dieter.bauer@uni-rostock.de}

\affiliation{Institut für Physik, Universität Rostock, 18051 Rostock, Germany}

\date{\today}

\begin{abstract}
Joint electron-ion energy spectra for the dissociative ionization of a model H$_2^+$ in few-cycle, infrared laser pulses are calculated via the numerical {\em ab initio} solution of the time-dependent Schrödinger equation. A strong, pulse-dependent modulation of the ionization probability for certain values of the protons' kinetic energy (but almost independent of the electron's energy) is observed. With the help of models with frozen ions, this feature---which mistakenly might be attributed to vibrational excitations---is traced back to the transient population of electronically excited states, followed by ionization. This assertion is further corroborated employing a two-level model incorporating strong-field ionization from the excited state.
\end{abstract}

%\keywords{}

\maketitle

\section{Introduction}
The hydrogen molecular ion  H$_2^+$ is one of the few systems for which the interaction with intense, short laser pulses can be simulated truly {\em ab initio}, i.e., based on the solution of the  time-dependent Schrödinger (TDSE) equation  \cite{Chelk_PhysRevA.52.2977,Hu_PhysRevA.80.023426} without further approximations such as, e.g., Born-Oppenheimer or Ehrenfest dynamics. Only rotations are usually neglected, which is justified for short laser pulses.  
Despite the simplicity of H$_2^+$, its joint electron spectra (JES) for electrons and nuclei are intriguingly complex
\cite{Madsen_PhysRevLett.109.163003,Silva_PhysRevLett.110.113001,Yue_PhysRevA.88.063420,Catoire_PhysRevA.89.023415}. In fact, on top of the already complex features in photoelectron spectra from atoms \cite{Milo_0953-4075-39-14-R01}   there is a nuclear degree of freedom added in H$_2^+$ (or its isotopic sisters). Hence, for any feature observed in a strong-field JES at least one question arises: are there vibronic excitations involved? 

Experimental photoelectron spectra for H$_2^+$  and JES for H$_2$  have been reported in Refs.~\cite{PhysRevA.89.013424,Wu_PhysRevLett.111.023002}, simulated ones in Refs.~\cite{Madsen_PhysRevLett.109.163003,Silva_PhysRevLett.110.113001,Yue_PhysRevA.88.063420,Catoire_PhysRevA.89.023415}. In the multiphoton regime, energy sharing according to $E_0 + n\hbar\omega = E_{e} + E_{p}$ is observed. Here, $E_0$ is the initial energy, $n\hbar\omega$ the absorbed photon energy, and $E_e$, $E_{p}$ the energy of the emitted electron and the nuclear kinetic energy release (KER), respectively. As $E_{e} + E_{p}=$ const., this correlated energy sharing leads to diagonal, straight-line features in the $E_e,E_{p}$-plane of the JES. At longer wavelengths the JES are less simple, especially at low electron energy where Coulomb effects are very important, as is well known from atomic strong-field ionization \cite{Wolter_1506.03636}. The diagonal, correlated features tend to fade while pronounced oscillations in the ionization probability as function of the electron energy emerge. However, also oscillations of the probability for dissociative ionization (DI) as function of the KER are observed, which have been shown to depend on the initial vibrational state \cite{Madsen_PhysRevLett.109.163003}.  One might be tempted to always attribute such variations in the DI probability to vibrational excitations. In fact, an interesting application of DI is Coulomb explosion imaging \cite{Chelk_PhysRevLett.82.3416}
where one strives for reconstructing the initial configuration of the nuclei from the KER spectrum after rapid ionization by a strong laser field. In this way, e.g., interference structures in the KER spectra due to a two-surface population dynamics in H$_2^+$ were observed experimentally \cite{staudte07,chelk07}. We will discuss  in this paper another mechanism that introduces a modulation in the KER. It is based on the oscillatory behavior of the ionization probability as function of the internuclear distance and the few-cycle laser pulse duration.

The paper is organized as follows. In Sec.~\ref{sec:correlated} we start with the full quantum H$_2^+$ model and introduce the effect we discuss in the remainder of this work: the ``vertical fringes'' (VF) in JES, indicating strong variations of the DI yield as function of the KER but almost independent of the electron energy. The subsequent sections serve to prove that the VF effect is {\em not} due to vibrational excitations (Sec.~\ref{sec:fixed-R}), {\em not} due to the one-dimensionality of our model (Sec.~\ref{sec:3Dfixed-R}), and {\em not} due to the two-center nature of diatomic molecular potentials (Sec.~\ref{sec:poetell}). In Sec.~\ref{sec:two-level}, a two-level model, combined with the strong-field approximation, is introduced that is capable of qualitatively reproducing the VF effect. We conclude in Sec.~\ref{sec:concl} and give all the numerical details in the Appendix, in particular on the t-SURFF approach for calculating the JES in the various geometries.

Atomic units $\hbar=m_{\text{e}}=|e|={4\pi \epsilon_0}=1$ are used unless otherwise indicated.

\section{Full quantum H$_2^+$ model}\label{sec:correlated}
\begin{figure}
\includegraphics[width=0.475\textwidth]{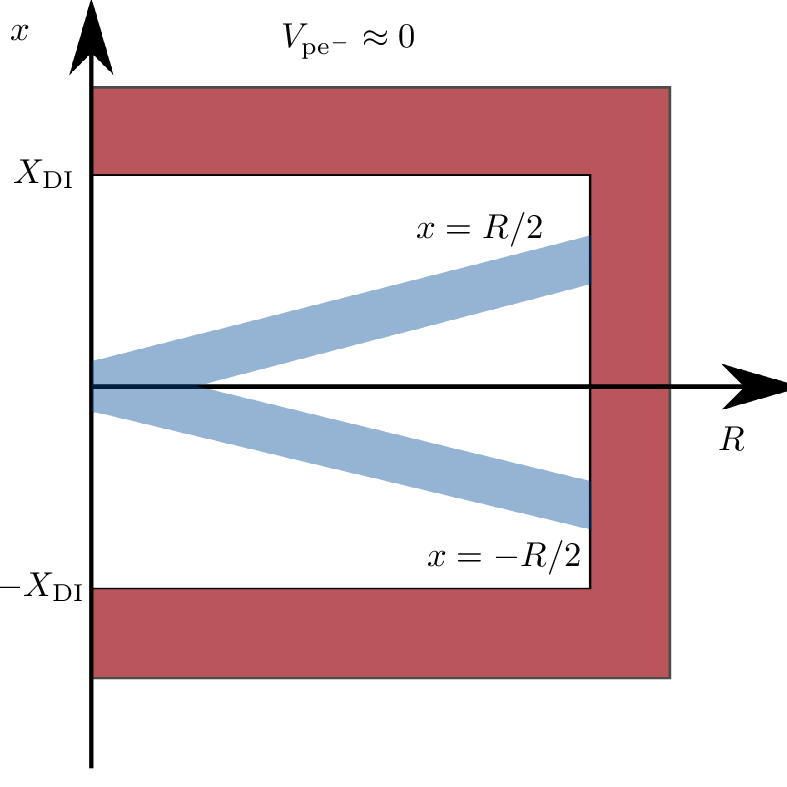}
\caption{\label{fig:proton_tsurff} (Color online)  Computational grid for the full quantum H$_2^+$ model. Coordinates $x$ and $R$ are electronic and internuclear distance, respectively.  
In the blue areas around $x=R/2$ and $x=-R/2$ the electron is close to one of the protons. The relevant t-SURFF boundary for DI (with electrons escaping in positive $x$ direction) is given by $x=X_{\text{DI}}$.
The red area  indicates the region in which a mask function absorbs probability density. }
\end{figure}

\begin{figure}
\includegraphics[width=0.475\textwidth]{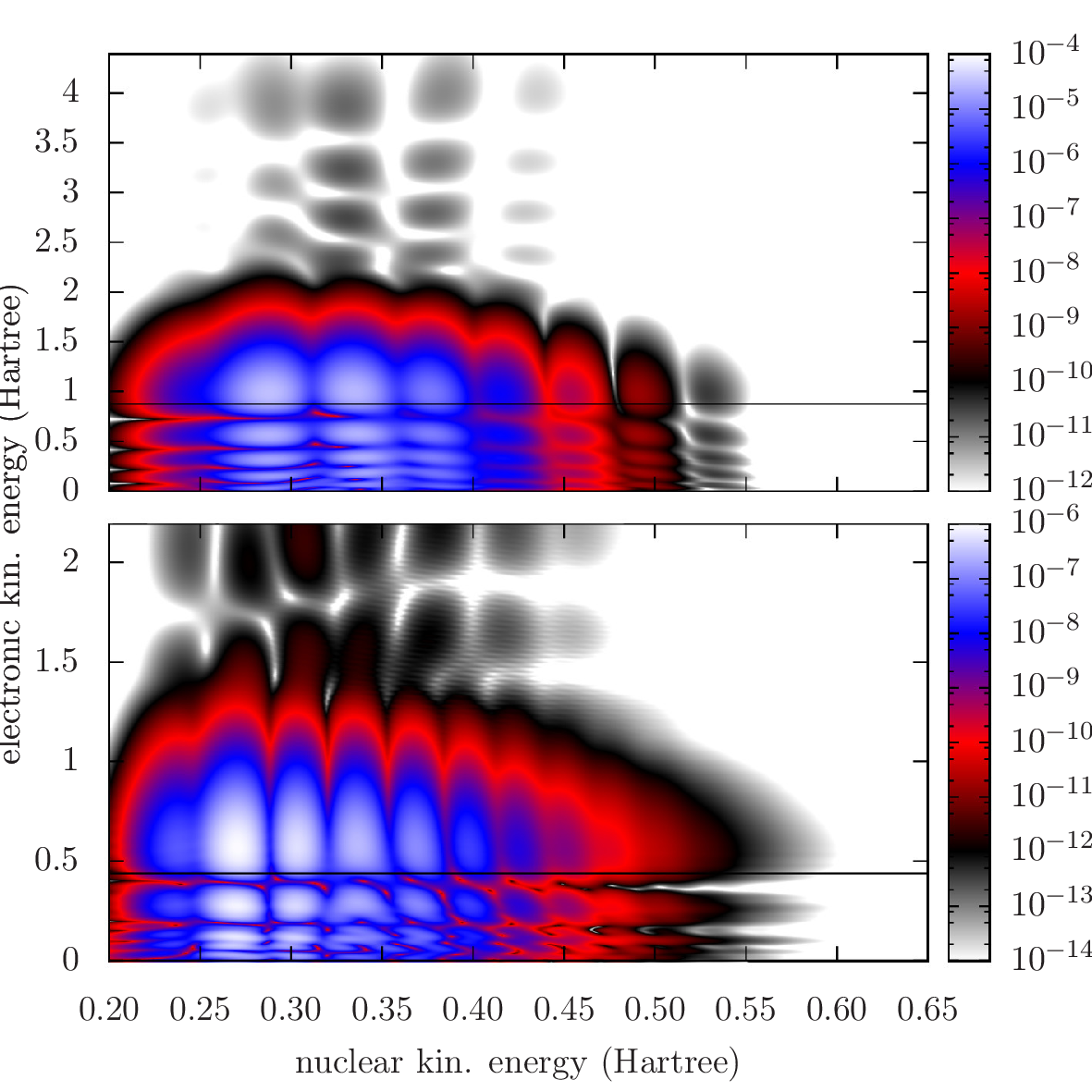}
\caption{\label{fig:correlated} (Color online)   Upper panel: JES for DI of H$_2^+$ in a $\sin^2$-shaped laser pulse with parameters: 
$n_{\text{c}}=3$, $\lambda=800$ nm, $I_{\text{peak}}=2.0\cdot 10^{14}$\,\Wcmcm, $U_p=0.44$. 
The horizontal line marks the cut-off for direct ionization ($2U_p$). The upper JES was calculated at the upper t-SURFF boundary $+X_\mathrm{DI}$.
Lower panel: $n_{\text{c}}=4$, $I_{\text{peak}}=1.0\cdot 10^{14}$\,\Wcmcm. The number of VF per nuclear kinetic energy interval increases with  $n_{\text{c}}$. The lower JES was calculated at the lower t-SURFF boundary $-X_\mathrm{DI}$.}
\end{figure}

The Hamiltonian for the two-dimensional H$_2^+$ model reads
\begin{equation}
  \label{eq:twod-hamiltonian}
  H=-\frac{1}{2\mu}\partial_x^2
  -\frac{1}{M}\partial_R^2
  -i\beta A(t)\partial_x 
  +V_{\text{pe}^-}
  +\frac{1}{|R|}\; .
\end{equation}
Both the electronic degrees and the nuclear degrees of freedom, i.e., electronic coordinate $x\in (-\infty,\infty)$ and internuclear distance  $R\in (0,\infty)$, are restricted to the laser polarization axis,  $M=1836$ is the proton mass, $\mu=2M/(2M+1)$, and $\beta=(M+1)/M$.  As we are dealing with a homonuclear diatomic molecule the laser field only couples to the electronic degree of freedom \cite{hiskes1961}. The velocity-gauge coupling to the laser field of vector potential $A(t)$ (in dipole approximation) was chosen, with the purely time-dependent $A^2$-term transformed away.
%We assume for all times the boundary condition $\Psi(x,R=0)=0$ for the solution of the TDSE,  
%which means that the exchange symmetry of the two protons has no effect on observables [DAZZLING SENTENCE ... VAGUELY REMEMBER WE DISCUSSED THAT BUT I FORGOT].

For the interaction between electron and protons we choose
\begin{equation}
  \label{eq:proton-electron-potential}
  V_{\text{pe}^-}=
  -\frac{1}{\sqrt{(x-R/2)^2+\epsilon}} 
  -\frac{1}{\sqrt{(x+R/2)^2+\epsilon}}
\end{equation}
with the smoothing parameter $\epsilon=1$.
% Somewhat unconventionally, the electronic  origin is located on one of the nuclei instead of the center of mass of the two nuclei as this simplifies the calculation of JES for DI via the  time-dependent surface flux method (t-SURFF) \cite{tao2012photo} (see Appendix~\ref{app:tsurff1} for details).   
The JES for DI is calculated via the  time-dependent surface flux method (t-SURFF) \cite{tao2012photo} (see Appendix~\ref{app:tsurff1} for details).
Figure~\ref{fig:proton_tsurff} depicts the geometry of the system.  Upon time-propagation, probability density will pass the surfaces defined by a sufficiently large $|x|=X_{\text{DI}}$, ``recorded'' there for the calculation of the JES using t-SURFF, and be absorbed by a mask function thereafter. 
For the t-SURFF approximation we have to assume $V_{\text{pe}^-}\approx 0$ which makes sense if $X_{\text{DI}}\gg R/2$.
% Electrons escaping in positive and negative $x$ direction can be separated are captured in this way. 
% If electrons escaping in the opposite direction are of interest as well one may run another simulation for $-A(t)$. 

% Results for full dynamic

% The ground state for the model Hamiltonian, obtained by the shift invert method, was used as initial state for the time propagtion.
Figure \ref{fig:correlated} shows correlated spectra for the process of DI in few-cycle    laser pulses of vector potential \begin{equation}
  \label{eq:vector_potential}
  A(t)=A_0\sin^2\left(\frac{\omega t}{2 n_{\mathrm{c}}}\right) \sin(\omega t)\end{equation}for $0 < t < T_{\text{p}}={2\pi n_{\mathrm{c}}}/{\omega}$,
calculated by the absolute square of (\ref{eq:DI_integral_final}). $\Tp$ is the pulse duration and $n_{\mathrm{c}}$ the number of laser cycles.
The initial wave function for the TDSE simulations was always the ground state  of the Hamiltonian (\ref{eq:twod-hamiltonian}), which has  the energy $E_{0}=-0.78$. The laser parameters are given in the figure caption.

The main features in the spectra in Fig.~\ref{fig:correlated} are nearly vertical and horizontal fringe patterns. Diagonal features indicating energy sharing according to $E_0 + n\hbar\omega = E_{e} + E_{p}$ are not observed for the laser parameters and the direction of escaping electrons chosen.
The modulation of the yield for fixed nuclear kinetic energy $E_p$ as function of the electronic kinetic energy $E_e$ is well known from laser atom interaction.
It can be attributed to the interference of several electron paths with different ionization times that lead to the same final electron momentum (see, e.g., \cite{lindner2005,Milo_0953-4075-39-14-R01,Arbo_PhysRevA.81.021403}).  Moreover, ``direct electrons'' and rescattered electrons can be clearly distinguished. The ``simple man's'' cut-off $2 U_p$ is indicated in both panels by a horizontal line. The yield due to direct electrons stretches well beyond $2 U_p$ before it drops down to the level of rescattered electrons (approximately four orders of magnitude smaller, visible for $0.26 < E_p < 0.47$).

The objective of this paper is to reveal the origin of the modulation of the yield as function of the nuclear kinetic energy $E_p$. 
In other words, why is the DI yield strongly suppressed for certain proton energies?
And why is this suppression almost independent of the electronic energy (i.e., why are the corresponding fringe patterns almost vertical in  Fig.~\ref{fig:correlated}) but dependent on the pulse duration?  Similar modulations have been reported in Ref.~\cite{Madsen_PhysRevLett.109.163003} for 
 simulations starting from a vibrationally excited H$_2^+$ molecule. The number of VF was found to  increase with increasing vibrational quantum number $\nu$ of the initial state. However, in our simulations we started from $\nu=0$ so that the vertical pattern in  Fig.~\ref{fig:correlated} does not just reflect the probability density of the initial vibrational wave packet. 

On one hand, experience shows that commonly all spectral features in strong field ionization can be explained in terms of interfering quantum trajectories. On the other hand, the interference of the usual long and short trajectories starting at the two nuclear sites \cite{lein2011} (including potential rescattering and the generation of a double-slit type interference pattern \cite{Spanner_0953-4075-37-12-L02}), should depend not only on the internuclear distance but also on the electronic energy $E_e$. Hence the strong suppression of the DI yield for certain values of $E_p$ but almost independent of $E_e$ cannot be explained by such interfering quantum trajectories.

\section{\label{sec:fixed-R} Fixed internuclear distance}
To rule out vibrations as the origin of the VF in Fig.~\ref{fig:correlated} results for H$_2^+$ with fixed internuclear distances are discussed now.

% results for molecular potential (fixed R).
\begin{figure}
\includegraphics[width=0.475\textwidth]{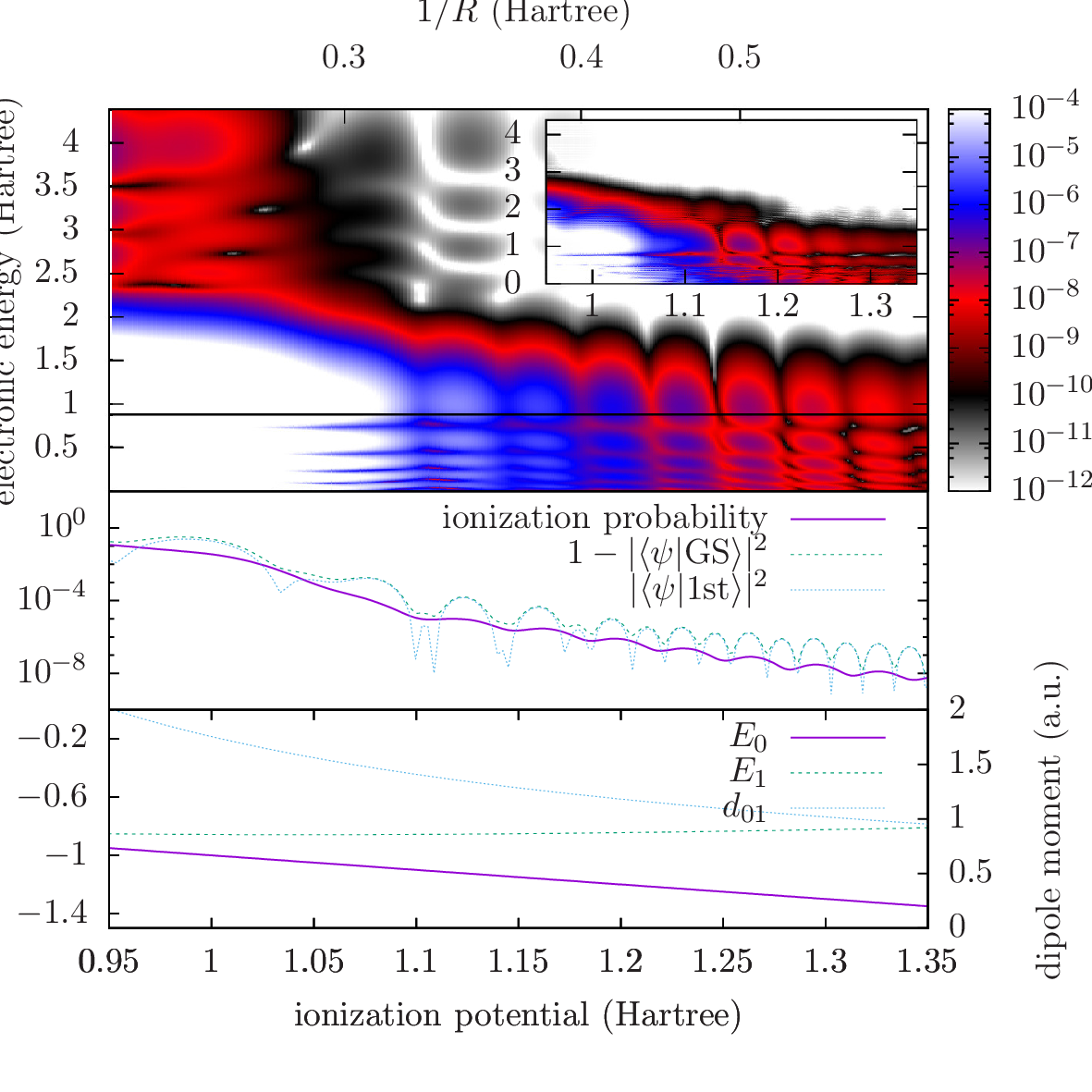}
\caption{\label{fig:oneD-fixed-R} (Color online)  Top panel: electronic spectra (each for fixed $R$)  for H$_2^+$ (same laser parameters as in Fig.~\ref{fig:correlated}, upper panel).
  Inlay in top panel: spectrum of electrons escaping in the opposite direction.
  Middle panel: occupation of the first excited state, $1-$ occupation of the ground state, and the total ionization probability.
  Bottom panel: energies  of ground and first excited state, and their transition dipole moment.
  }
\end{figure}

In order to calculate electronic spectra for the H$_2^+$ model with fixed inter nuclear distance $R$ the electronic TDSE $ i \partial_t \Psi(x,t) = H(t) \Psi(x,t)$ for the Hamiltonian
\begin{equation}
  \label{eq:oned-hamiltonian}
  H(t)=-\frac{1}{2}\partial_x^2-i A(t)\partial_x + V(x)
\end{equation}
and binding potential
\begin{equation}
  \label{eq:oned-mol-pot}
  V(x)= -\frac{1}{\sqrt{(x-R/2)^2+\epsilon}} -\frac{1}{\sqrt{(x+R/2)^2+\epsilon}}
\end{equation}
was solved for many $R$ in the range where $1/R \simeq E_p$ covers the relevant KER $E_p$  in Fig.~\ref{fig:correlated}.

In the top panel of Fig.~\ref{fig:oneD-fixed-R} all these electronic spectra are collected for comparison with Fig.~\ref{fig:correlated}, upper panel. 
The overall trend is an increasing ionization yield with increasing $R$ because of the decreasing ionization potential $I_p=|E_0|$ (see bottom panel).
The most important insight gained from these fixed-$R$ simulations is that pronounced suppressions of the ionization yield are observed for certain internuclear distances $R\simeq 1/E_p$ as well. This proves that vibrational excitation cannot be the origin of the VF visible in both Fig.~\ref{fig:oneD-fixed-R} and Fig.~\ref{fig:correlated}. 

The bottom panel of Fig.~\ref{fig:oneD-fixed-R} shows the energies of the two lowest bound electronic states in the H$_2^+$ potential \reff{eq:oned-mol-pot} vs the ionization potential $I_p=|E_0|$.
For large internuclear distances $R$ the two levels are almost degenerate with the ground state energy rising asymtotically towards the ground state energy value for the potential $V(x)=-1/\sqrt{x^2+\epsilon}$. Because of this asymptotic degeneracy and the related diverging transition dipole moment $d_{01}$  these two states were dubbed ``charge resonance states''. However, we should stress that the VF in the DI yield as discussed in this work occur at smaller distances than ``charge resonance enhanced ionization'' (CREI) \cite{CREI_PhysRevA.52.R2511}.

The middle panel of Fig.~\ref{fig:oneD-fixed-R} shows the total ionization probability
\begin{equation}
  \label{eq:total-ion-prob}
  P_{\text{ion}}=1-\int_{-X_{\text{I}}}^{X_{\text{I}}} \diff x\, |\Psi(x,T)|^2 \; ,
\end{equation}
and the occupations of the ground and first excited states after the interaction with the laser pulse.
The modulations in the total ionization probability is less pronounced than in the energy resolved spectrum, which can be explained by the ``left-right asymmetry'' \cite{Milo_0953-4075-39-14-R01} of the spectra for electrons escaping in polarization direction (top panel) and  opposite to it (inset in top panel). 
Moreover, the modulations in the energy resolved spectrum are not strictly independent of the electronic energy $E_e$, i.e., not perfectly vertical but slightly tilted.

The fact that ionization probability and bound state occupations at the end of the pulse oscillate similarly as function of $I_p$ (or $1/R\simeq E_p$) suggests that the electronically excited state plays an important role in the DI process in few-cycle laser pulses. However, the minima in the ionization yield do not perfectly coincide with the minima in the excited-state population. For ionization probabilities smaller than $10^{-6}$ the occupation of the excited state rather oscillates with twice the frequency of $P_{\text{ion}}$ as function of $I_p$. Hence, the
 ionization step introduces an additional, nontrivial $I_p$-dependence. Note that the excited state energy  $|E_1|\simeq 0.8$ varies little with $R$ so that in any case at least about $12$ photons are required for ionization. Additionally, 2 up to 10 photons are needed to couple the initial electronic ground state of energy $E_0$ with the excited state of energy $E_1$. Below, in Sec. \ref{sec:two-level}, we reproduce the VF qualitatively, using a two-level approximation in combination with the strong-field approximation (SFA).

\section{\label{sec:3Dfixed-R} 3D H$_2^+$ with fixed internuclear distance}
In order to show that the modulation in the DI yield  is not an artifact of the low dimensionality of our models  the 3D molecular ion with fixed ions and aligned along the laser polarization axis was considered.
 The Hamiltonian
\begin{multline}
  \label{eq:cylind_hamil}
  H=\frac{1}{2}\left( -i\pmb{\nabla}+A(t)\pmb{e}_z\right)^2\\ 
  -\frac{1}{\sqrt{(z-R/2)^2+\rho^2}} -\frac{1}{\sqrt{(z+R/2)^2+\rho^2}}
\end{multline}
is cylindrically symmetric so that
 the natural choice for the t-SURFF boundary is the surface of a cylinder with radius $R_{\text{I}}$ and height $2 Z_{\text{I}}$ (see Appendix \ref{sec:app3dcyl} for details).

Figure \ref{threeD-fixed-R} shows spectra for various internuclear distance $R$ and electrons escaping in polarization direction. The modulation of the ionization yield  as function of $1/R$ is clearly visible
although at low electron energies the fringes are more tilted than in the 1D results, making the suppression of the yield for certain internuclear distances less electron energy-independent. The fringe pattern for the rescattered electrons instead is as vertical as in the 1D results. Revealing the origin of this difference between 1D and 3D results requires further systematic investigations. In this work we are content with the fact that the modulation in the ionization yield as function of $1/R$ exists in 3D as well.

\begin{figure}
\includegraphics[width=0.475\textwidth]{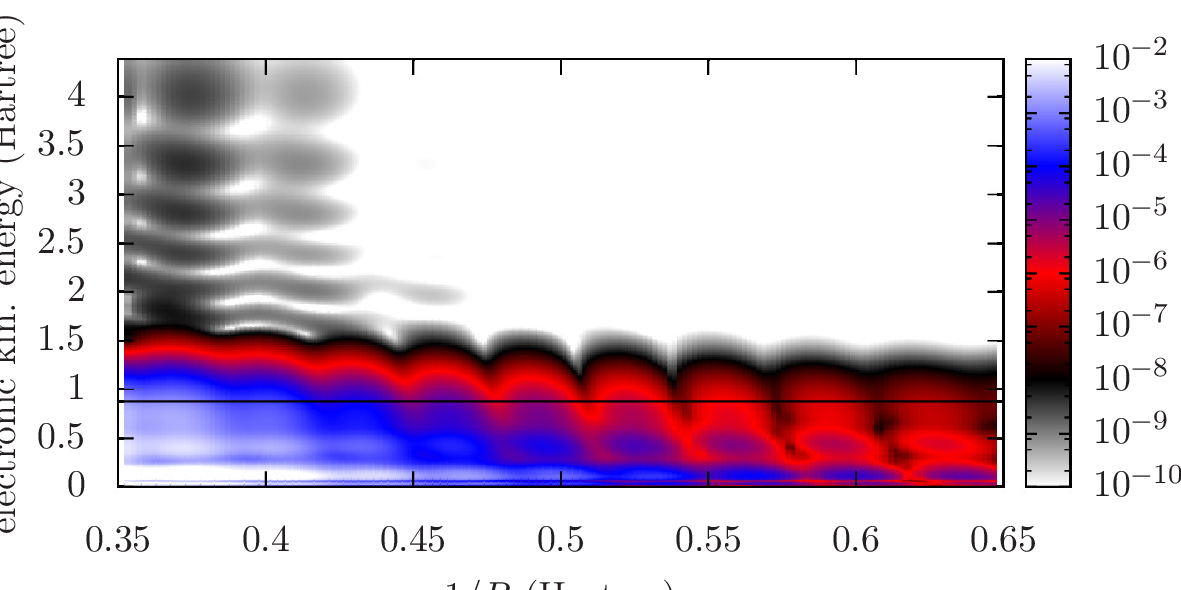}
\caption{\label{threeD-fixed-R} (Color online)  Same as Fig.~\ref{fig:oneD-fixed-R} but for 3D H$_2^+$ aligned in polarization direction of the laser.}
\end{figure}

\section{Single-center potential} \label{sec:poetell}
Next we show that the two-center nature of the binding potential is not essential for the observed modulations of the ionization yield, while the existence of an excited state is. 
To that end we consider a Pöschl-Teller potential of the form
\begin{equation}
  \label{eq:oned-single-pot}
  % V(x)=-\frac{2}{\sqrt{\epsilon+x^2}}
  V(x)=-\frac{b(b-1)}{8 a^2 \cosh^2[x/(2 a)]} 
\end{equation}
for which the finite number of energy levels of energy $E_n=-(b-n-1)^2/(8a^2)$, $n=0,1,2, \ldots < b-1$ can be adjusted via the parameters $a>0$ and $b>1$.

% Poeschl Teller simultion for several boud states
\begin{figure}
\includegraphics[width=0.475\textwidth]{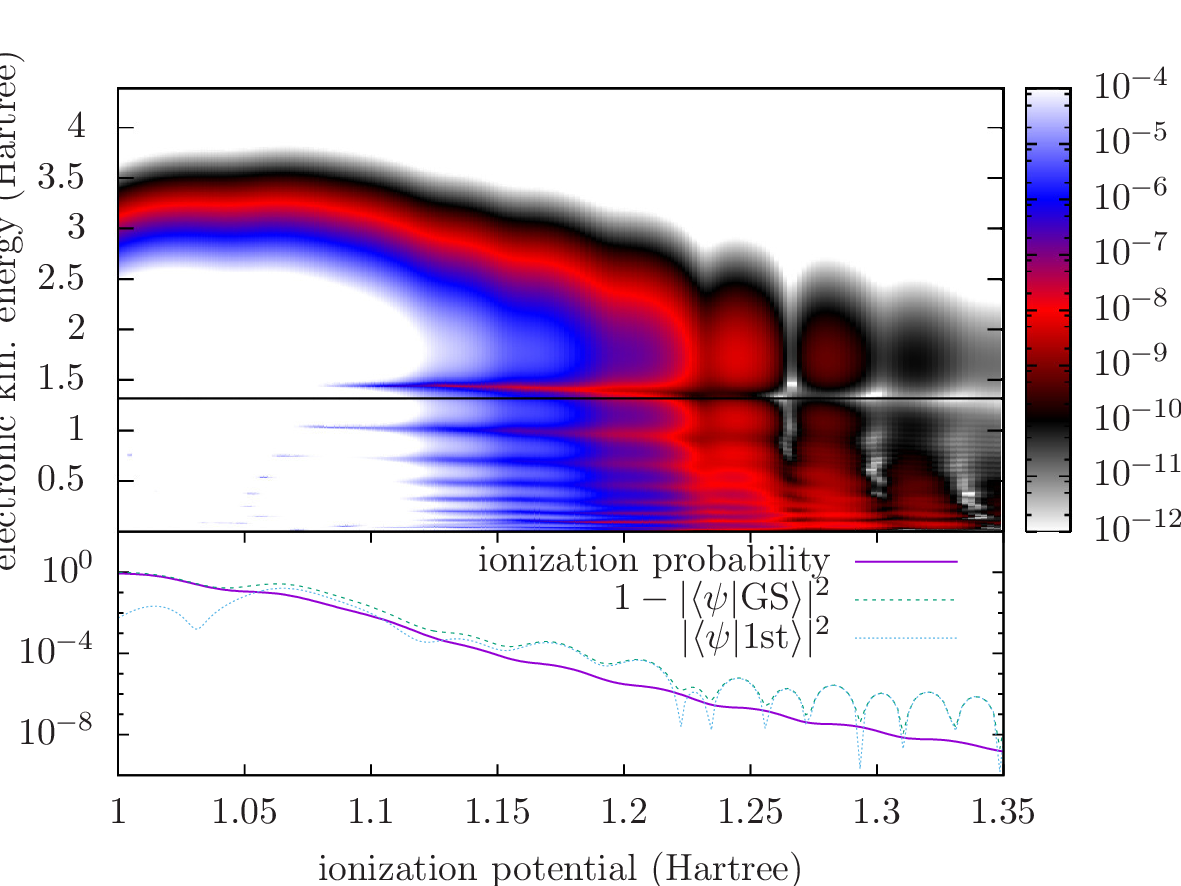}
\caption{\label{poeschl-teller-multi} (Color online) 
Same as Fig.~\ref{fig:oneD-fixed-R} but for a Pöschl-Teller potential \eqref{eq:oned-single-pot} with ground and excited state energy tuned close to the molecular case. The laser intensity was increased to $I_{\text{peak}}=3.0\cdot 10^{14}$\,\Wcmcm\ in order to have an ionization yield similar to the molecular models.}
\end{figure}

First, we aim at mimicking the behavior of ground and excited state in the molecular model, i.e.,  $E_1=-0.85$ is kept constant, and $E_0$ covers the range $-1.35<E_0<-1$. Figure \ref{poeschl-teller-multi} shows the electron spectra collected such that they can be directly compared to Fig.~\ref{fig:oneD-fixed-R}.
The VF are there, proving that they are not due to a two-center interference.

\begin{figure}
  \includegraphics[width=0.475\textwidth]{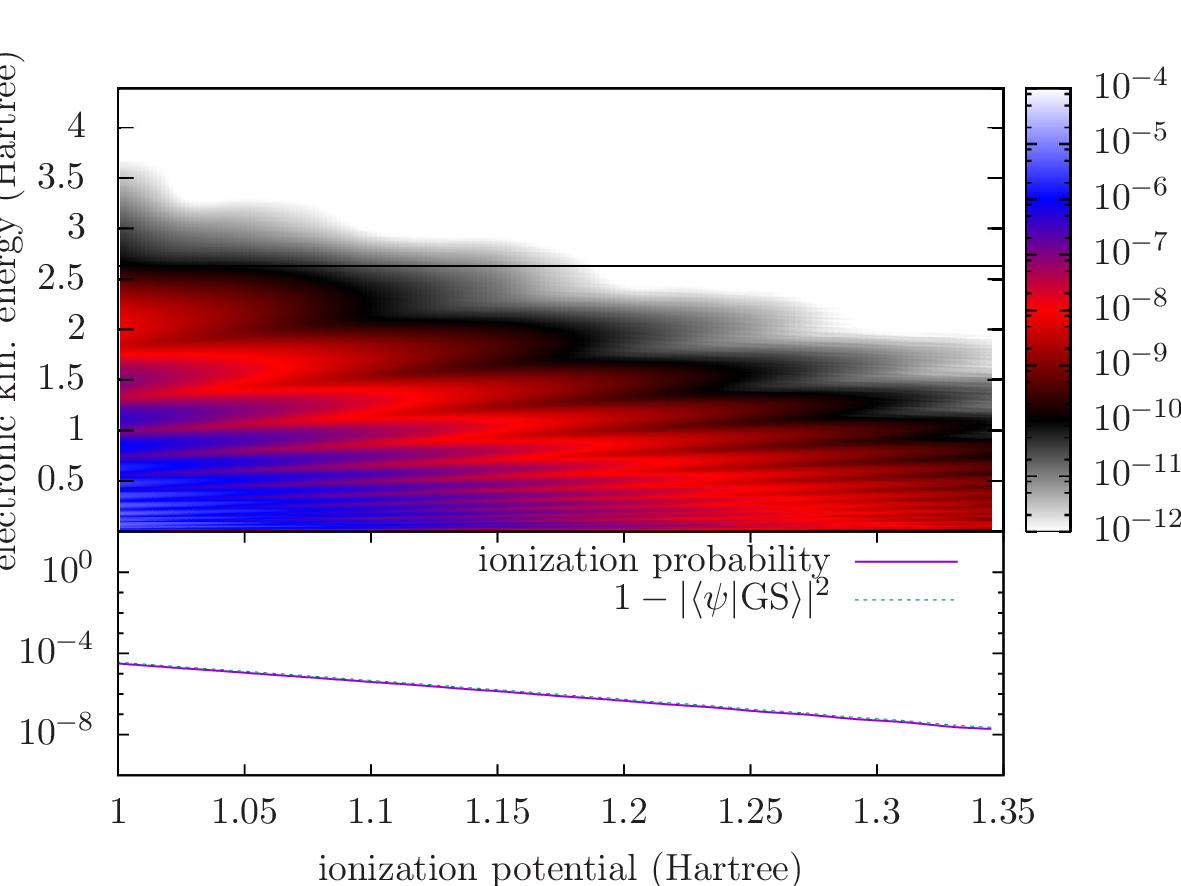}
\caption{\label{poeschl-teller-single} (Color online) 
Same as Fig.~\ref{poeschl-teller-multi} but for a Pöschl-Teller potential \eqref{eq:oned-single-pot}   supporting only a single bound state. The intensity was increased further to $I_{\text{peak}}=6.0\cdot 10^{14}$\,\Wcmcm.}
\end{figure}

Second, Fig.~\ref{poeschl-teller-single} shows the case of a Pöschl-Teller potential with a single bound state only.
The intensity was increased to $I_{\text{peak}}=6.0\cdot 10^{14}$\,\Wcmcm\ to compensate for the decreasing ionization in the narrower and deeper potential. Each individual photoelectron spectrum looks standard ``SFA-like''.
Both the VF and oscillations in the occupation of the groundstate at the end of the laser pulse are absent.
This substantiates our assertion that the occupation of an excited state is crucial for the modulation of the (dissociative) ionization yield.

% result for a Poeschl Teller potential in one dimension with only one bound stae

\section{\label{sec:two-level}Two levels + SFA}
\begin{figure}
  \includegraphics[width=0.475\textwidth]{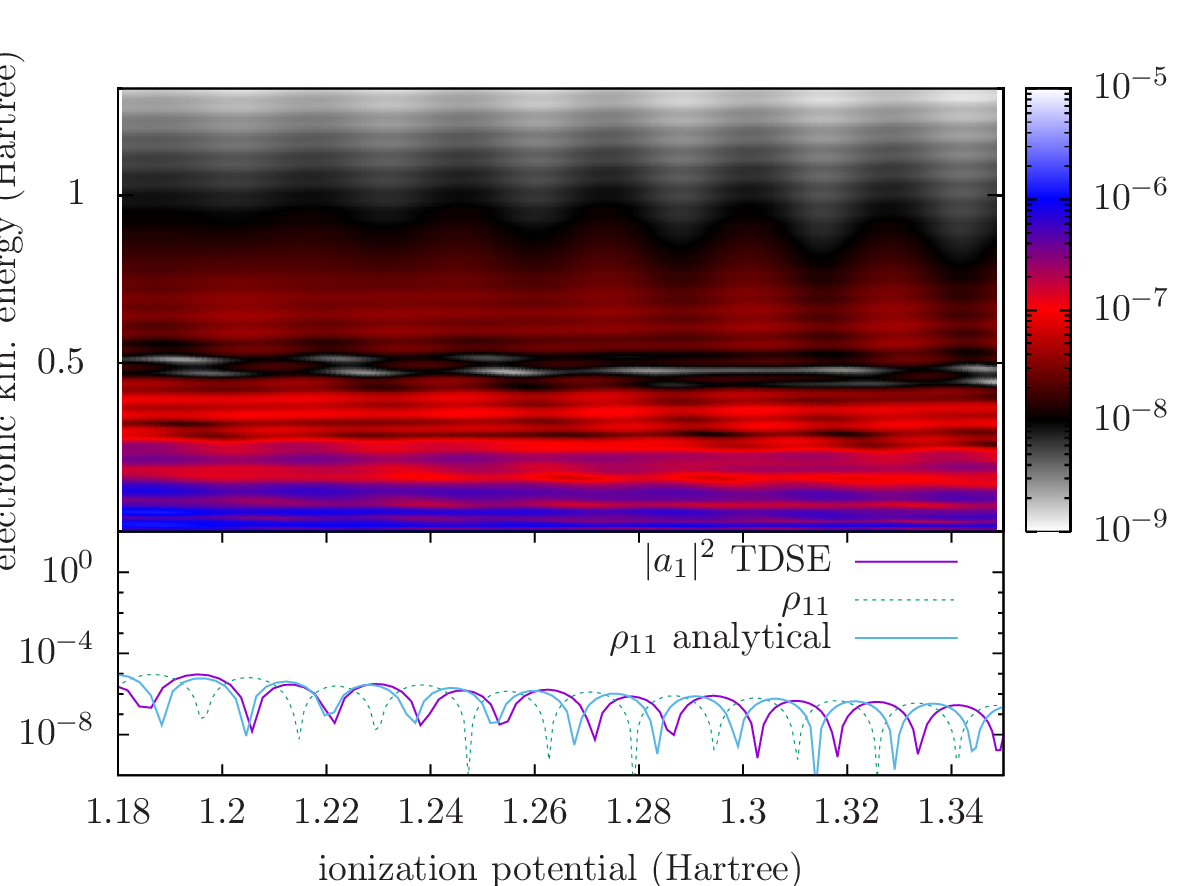}
\caption{\label{fig:sfa-two-level} (Color online) 
Upper panel: photoelectron spectra for the model of Fig.~\ref{fig:oneD-fixed-R} calculated using the two-level SFA.
Lower panel: occupation of the first excited state as calculated from the TDSE, the numerical solution of  Eqs.~(\ref{eq:coupled-dmeq-00}) and (\ref{eq:coupled-dmeq-10}), and the analytical result \reff{eq:analytical-rho}. 
}
\end{figure}

As long as the ionization probability is small
we may model the occupation of the ground and first excited states by a simple two-level model.
Plugging the ansatz
$\ket{\psi(t)} =  a_0(t) \ket{\psi_0} + a_1(t) \ket{\psi_1} $
into the TDSE in length gauge
\begin{equation}
\imagi\partial_t  \psi(x,t) = \left(-\frac{1}{2}\partial_x^2 + E(t)x + V(x) \right) \psi(x,t)   
\end{equation}
one finds the well-known equations of motion for the density matrix elements $\rho_{ij}=a_i^*a_j$, $i=0,1$, $j=0,1$, 
\begin{align}
  \label{eq:coupled-dmeq-00}
  \dot{\rho}_{00} &= \imagi d_{01} E(t)(\rho_{10}-\rho_{10}^*)\\
  \label{eq:coupled-dmeq-10}
  \dot{\rho}_{10} &= \imagi d_{01} E(t)(\rho_{00}-\rho_{11})+\imagi \Delta E \rho_{10} 
\end{align}
where $\Delta E=E_1-E_0$, $d_{01}=\langle\psi_0|x|\psi_1\rangle$ (assumed real), $\rho_{01}=\rho^*_{10}$, $\rho_{11}=1-\rho_{00}$. As we are interested in few-cycle pulses and the transient dynamics induced by them we cannot apply the rotating wave approximation, and a dressed or Floquet state approach does not make sense either.
Instead, in the bottom panel of Fig.~\ref{fig:sfa-two-level} the density matrix element $\rho_{11}$ at the end of the laser pulse $t=\Tp$ from the numerical solution of the two-level model Eqs.~\eqref{eq:coupled-dmeq-00} and \eqref{eq:coupled-dmeq-10} [initial conditions  $\rho_{00}(t=0)=1$ and $\rho_{11}(t=0)=0$] is compared 
to the occupation of the first excited state from the numerical solution of the  full TDSE.
The agreement is very good apart from a  shift along the ionization potential axis.
This shift is caused by neglecting higher excited states and the coupling to the continuum in the two-level model. We checked that for lower field strengths (where even less ionization occurs and other excited states are negligibly populated) the agreement improves.

% occupation of the first excited state at end of the pulse

The results of the two-level system can be used to model ionization as well. In ``standard'' SFA only a single bound state (besides the continuum states of momentum $k$) is considered, and depletion of its population is neglected (see, e.g., Ref.~\cite{Milo_0953-4075-39-14-R01}). 
Instead, we plug  the modulus of the occupation $|a_1(t)|=\sqrt{\rho_{11}(t)}$ of the first excited state into the SFA amplitude for direct ionization,
\begin{widetext}
%\begin{multline} 
\beq\lvert a_{\text{I, SFA}}(k,t) \rvert^2 
%=
% \Bigg\vert\int^t \! \diff t'\, \sqrt{\rho_{00}(t)} d_{0k}(A(t') + k) E(t')  \eulere^{ \imagi \int_{0}^{t'} \! \diff t''\,  \frac{[A(t'') + k]^2}{2}-E_0}\Bigg\vert^2\\  +
\simeq \Bigg\vert\int^t \! \diff t'\, \sqrt{\rho_{11}(t)} d_{1k}(A(t') + k) E(t')  \eulere^{ \imagi \int_{0}^{t'} \! \diff t''\,  \frac{[A(t'') + k]^2}{2}-E_1}\Bigg\vert^2 \; .
%\end{multline}
\eeq
Here, $d_{1k}(k)=\bra{\psi_1} x \ket{k}$, and we neglect the amplitude for ionization from the ground state because its contribtion is several orders of magnitude smaller than the contribution of the first excited state.
This is fortunate, as otherwise the phases of the complex $a_0(t)$ and $a_1(t)$ matter and should be calculated from a full SFA with two bound states and all relevant bound-bound and bound-continuum couplings. 
The top panel in Fig.~\ref{fig:sfa-two-level} shows the collected electronic spectra for the molecular potential with fixed protons of Sec.~\ref{sec:fixed-R}, calculated using our simplified  two-level SFA.
The dipole moments and the eigenenergies $E_0$ and $E_1$ were calculated numerically from the TDSE data of Sec.~\ref{sec:fixed-R}  (see the bottom panel of Fig.~\ref{fig:oneD-fixed-R} for the $R$-dependent energies and transition dipole moment $d_{01}$). Figure~\ref{fig:sfa-two-level} shows that the simple two-level SFA reproduces the $I_p$-dependent features in the ionization probability qualitatively. In particular, the correlation between the oscillations of the excited-state population at the end of the pulse as function of $I_p$  and the oscillations in the ionization probability with only half the frequency is as observed in the TDSE results of Sec.~\ref{sec:fixed-R}.

Employing $\rho_{11}\ll \rho_{00} = 1- \rho_{11}\simeq 1$ in (\ref{eq:coupled-dmeq-10}), Eqs.~(\ref{eq:coupled-dmeq-00}) and (\ref{eq:coupled-dmeq-10}) can be solved analytically, leading to
\beq \rho_{11}(\tau) = \left| \frac{d_{01}}{\omega} \int^\tau\diff \tau'\,E(\tau') \eulere^{-\imagi n  \tau' } \right|^2, \qquad \tau=\omega t, \quad \ n=\Delta E/\omega . \label{rho11} \eeq
Using $E(t)=-\partial_t A(t)$ with the vector potential (\ref{eq:vector_potential}) yields  the occupation of the first excited state at the end of the $\sin^2$ pulse 
%\begin{multline}
\beq  \label{eq:analytical-rho}
  \rho_{11}=
  \left[ \frac{A_0 d_{01} \omega^3 (n_{\text{c}}^2 \omega^2-\omega^2+3 \Delta E^2 n_{\text{c}}^2)  \Delta E \sin\left(\pi \frac{\Delta E}{\omega} n_{\text{c}}\right)}
  {\omega_{-} \omega_{+} (n_{\text{c}} \omega_{-} \! -\omega) (n_{\text{c}} \omega_{+} \! -\omega) (n_{\text{c}} \omega_{-} \! +\omega) (n_{\text{c}} \omega_{+} \! +\omega)} \right]^2
%\end{multline}
\eeq
\end{widetext}
where $\omega_{+}=\omega+\Delta E$ and $\omega_{+}=\omega-\Delta E$.
Expanding this expression in the small parameter $\eta=\omega/\Delta E=1/n$ gives 
\begin{equation}
  \label{eq:rho-expansion}
  \rho_{11} \simeq 
  \frac{9 A_0^2 d_{01}^2 \sin^2(\pi \Delta E n_{\text{c}}/\omega) \eta^6}{n_{\text{c}}^4} .
\end{equation}
The occupation of the first excited state after the pulse thus decreases as $n_{\text{c}}$ increases. Hence the observed VF in the (dissociative) ionization yield are a few-cycle effect. Moreover, inspection of the sine's argument in (\ref{eq:analytical-rho}) shows that the frequency of the oscillation depends on the number of cycles $n_{\text{c}}$. The higher $n_{\text{c}}$ the more oscillations within a given $\Delta E/\omega$ interval. This is in agreement with the TDSE results in Fig.~\ref{fig:correlated} where  the 4-cycle laser pulse was found to generate more VF than the $3$-cycle pulse. As $E_1$ is almost constant in the H$_2^+$ model the oscillations in Fig.~\ref{fig:sfa-two-level} are of almost constant period when plotted vs $I_p=|E_0|$.

Note that $\rho_{11}$ is very sensitive to the pulse shape. In fact, for a Gaussian pulse $\rho_{11}$ according \reff{rho11}---with the integration limits stretched to $\pm\infty$---becomes the (modulus squared) Fourier transform of a Gaussian, which is a Gaussian and thus does not oscillate with  $\Delta E$ and $n_{\text{c}}$ (full-width half maximum) at all.

\section{Conclusion} \label{sec:concl}
Numerical simulations of the dissociative ionization  process in H$_2^+$ for short laser pulses reveal patterns of vertical fringes in the joint energy spectra, i.e., strong variations of the yield as function of the ion energy that are almost independent of the electron energy.
Identifying the kinetic energy release with the inverse internuclear distance, the effect is also found in calculations with fixed ions, ruling out vibrational excitations as its origin. Instead, ionization proceeds via the first excited electronic state.  In few-cycle pulses the population of the first excited state depends strongly on the number of cycles and the pulse shape in general. The vertical fringes in the continuous dissociative ionization spectra are clearly correlated with the population of the  first excited state at the end of the pulse, as qualitatively reproduced using a simple two-state model combined with the strong-field approximation. 

The observed effect relies on the ultrashort, transient dynamics in few-cycle laser pulses and {\em not} on resonances, specially chosen  detunings, or interference. In fact, in the limit of long laser pulses the vertical fringes disappear and one approaches---depending on the laser frequency---either ordinary non-resonant multiphoton or tunneling ionization, or well-known resonance-enhanced multiphoton ionization.

%\section*{Acknowledgment}
%This work was supported by the SFB 652 of the German Science Foundation (DFG).

\begin{acknowledgments}
This work was supported by the SFB 652 of the German Science Foundation (DFG).
\end{acknowledgments}

\appendix
% \section{Correlated spectra for DI in H$_2^+$}

\section{Numerical details} \label{app:numeric}
The TDSE was solved numerically by propagating the wavefunction with the Crank-Nicolson time propagator.
The wavefunction and the potentials were discretized on a Cartesian grid with
 the spatial derivatives in the Hamiltonian approximated by finite differences.
  An iterative block Gauss-Seidel method and  the Thomas algorithm were applied for the solution of the linear system of equations of the Crank-Nicolson method in the two and one dimensional case, respectively.
The initial ground-state wavefunctions for the time propagation and the first excited state were  obtained by the shift-invert method \cite{saadev}.
For the numerical solution of the cylindrically symmetric Hamiltonian (\ref{eq:cylind_hamil}) the coordinate transformation $\xi=\rho^{3/2}$ was used \cite{kono}.
Numerical parameters for the TDSE simulations are summarized in Table~\ref{tab:parameters}.

 \begin{widetext}

\section{t-SURFF for the H$_2^+$ model} \label{app:tsurff1}
Assuming that $V_{\text{pe}^-}$ in  \eqref{eq:proton-electron-potential} can be neglected for $x>X_{\text{DI}}$ the wavefunction there separates in the form  
$\psi_k(x,t) \phi_p(R)\eulere^{-\imagi t E_p}$ where $\phi_p(R)$ are the solutions of the Coulomb scattering problem
\begin{equation}
  \label{eq:coulomb_scattering}
  \left(T_{\text{N}}+\frac{1}{R}\right)\phi_p(R)=E_p \phi_p(R) \; , \qquad T_{\text{N}}=-\frac{1}{M} \partial_R^2,  
\end{equation}
and
\beq
  \label{eq:volkov_oneD}
  \psi_{k}(x,t)=(2\pi)^{-1/2} \eulere^{-\imagi \alpha(t)+\imagi k x},\qquad \alpha(t) = \frac{1}{2}\int_0^t \!d t'\, [k^2+2kA(t')]
\eeq
are Volkov wavefunctions.

The DI amplitude  (restricted to the electrons escaping in positive direction) is approximated by the integral
%\begin{align}
%\begin{multline}
 \beq \label{eq:DI_probability_amplitude}
  \langle \Psi(T_{\text{p}})|DI \rangle
  \simeq a_{\text{DI}}(k,p)\equiv 
  \langle \Psi(T)|\Theta(x-X_{\text{DI}}) |p(T) \rangle |k(T) \rangle 
  =\int \! d R \! \int\limits_{x>X_{\text{DI}}} \!d x\; \Psi^*(x,R,T) \psi_k(x,T) \phi_p(R)\eulere^{-\imagi T E_p} \; .
%\end{multline}
% \end{align}
\eeq

This expression is not yet useful for practical purposes because $T$ needs to be large enough to allow the slow electrons arriving in the region $x>X_{\text{DI}}$. On the other hand, the fast electrons need to be kept on the grid as well,  necessitating a huge grid size.
In order to avoid large grids the t-SURFF method \cite{tao2012photo,Yue_PhysRevA.88.063420} was adapted to the problem at hand.
Writing the right hand side of (\ref{eq:DI_probability_amplitude}) as a time integral we obtain
\beq
%\begin{multline}
  a_{\text{DI}}(k,p) = \langle \Psi(0)|\Theta(x-X_{\text{DI}}) |p(0) \rangle |k(0) \rangle 
  + \int_0^T\!d t\, \partial_t \langle \Psi(t)|\Theta(x-X_{\text{DI}}) |p(t) \rangle |k(t) \rangle  \; .\label{eq:DI_time_integral}
%\end{multline}
\eeq
For sufficiently large $X_{\text{DI}}$ and a bound initial state $|\Psi\rangle$ we have  $\langle \Psi(0)|\Theta(x-X_{\text{DI}}) |p(0) \rangle |k(0) \rangle\simeq 0$. Employing the TDSE with $V_{\text{pe}^-}\simeq 0$ yields
\beq
%\begin{multline}
  \label{eq:DI_time_integral_commutator}
  % \langle \Psi(T)|\Theta(x-X_{\text{DI}}) |p(T) \rangle |k(T) \rangle 
  a_{\text{DI}}(k,p)\simeq 
  a_{\text{DI,t-SURFF}}(k,p) 
  \equiv \imagi \int_0^T \!d t \, \langle \Psi(t)|\Big[-\frac{1}{2\mu}\partial_x^2
  -\frac{1}{M}\partial_R^2
  -\imagi\beta A(t)\partial_x
  +\frac{1}{R},\Theta(x-X_{\text{DI}}) \Big] |p(t) \rangle |k(t) \rangle \; .
%\end{multline}
\eeq
Only terms of the Hamiltonian containing derivatives with respect to $x$ contribute in the commutator, leading to 
\begin{multline}
  \label{eq:DI_integral_final}
  % \langle \Psi(T)|\Theta |p(T) \rangle |k(T) \rangle 
  a_{\text{DI,t-SURFF}}(k,p) 
  =\int_0^T \!d t \int_0^{\infty}\!d R \Big[\beta A (t)\Psi^*(X_{\text{DI}},R,t)\psi_k(X_{\text{DI}},t) \\
  -\frac{\imagi}{2\mu}\big(\Psi^*(X_{\text{DI}},R,t)\partial_x\psi_k(x,t)|_{x=X_{\text{DI}}} 
    -\psi_k(X_{\text{DI}},t)\partial_x\Psi^*(x,R,t)|_{x=X_{\text{DI}}}\big) \Big] \phi_p(R,t) .
\end{multline}
The scattering states $\phi_p(R)$ were used as implemented in the GNU Scientific Library (GSL) \cite{gsl}.

% The wavefunctions obtained in the TDSE simulations are not symmetric with respect to exchange of the two protons.
% In practice this does not pose a problem since the repulsive potential $V_{\text{pp}}=1/R$ renders proton exchange unlikely in the energy range relevant for the results discussed below.

% Integral \ref{eq:DI_integral_final} was approximated by the trapezoidal rule. 
% The grid size  in the direction$R$

\section{t-SURFF for 1D calculations} \label{app:tsurff1D}
For the one dimensional systems the probability amplitude for ionization with final electron momentum $k$ is approximated as
\begin{multline}
  \label{eq:ion_amplitude_oneD}
  \langle \Psi(T)|\Theta(x-X_{\text{I}})  |k(T) \rangle 
  \simeq \int_0^T \!d t 
  \Big(A(t)\Psi^*(X_{\text{I}},t)\psi_k(X_{\text{I}},t)
  -\frac{i}{2}\left(\Psi^*(X_{\text{I}},t)\partial_x\psi_k(x,t)|_{x=X_{\text{I}}}
    -\psi_k(X_{\text{I}},t)\partial_x\Psi^*(x,t)|_{x=X_{\text{I}}}\right) \Big).
\end{multline}
Again, only electrons escaping in positive $x$ direction, passing the t-SURFF boundary $X_{\text{I}}$, are considered, 
and the binding potential is neglected for distances $x>X_{\text{I}}$.

\section{t-SURFF for cylindrically symmetric system} \label{sec:app3dcyl}
The probability amplitude for an electron escaping with a momentum $\pmb{k}=k_{\rho}\pmb{e}_x+k_{z}\pmb{e}_z$ can be approximated by the integral
\beq
%\begin{multline}
  \label{eq:cylind_int}
  \langle\pmb{k}(T)|\Psi(T)\rangle 
  \simeq \int_0^T \!dt\int\!dV \partial_t\Big(\psi_{\pmb{k}}(\rho,z,t)^* \big(\theta(R_{\text{I}}-\rho)\theta(z-Z_{\text{I}})
  +\theta(R_{\text{I}}-\rho)\theta(-z-Z_{\text{I}})+\theta(\rho-R_{\text{I}}) \big) \Psi(\rho,z,t) \Big)\; .
%\end{multline}
\eeq
The integral (\ref{eq:cylind_int}) can be divided into three terms which are evaluated separately.
Using the TDSE, the first term reads (dropping the arguments of $\psi_{\pmb{k}}$ and $\Psi$)
\begin{eqnarray}
 \label{eq:summand_1}
  s_1(k_{\rho},k_z) &=& \int_0^T \!dt\int\!dV \partial_t\Big(\psi_{\pmb{k}}^* \theta(R_{\text{I}}-\rho)\theta(z-Z_{\text{I}})  \Psi \Big) \nonumber\\
  &=& -\frac{i}{2} \int_0^T \!dt\int dV \psi_{\pmb{k}}^* \left[\rho^{-1} \partial_{\rho}\rho\partial_{\rho} + \partial_z^2 + 2 iA(t) \partial_{z}, \theta(R_{\text{I}}-\rho)\theta(z-Z_{\text{I}})  \right]\Psi\nonumber\\
&=& \frac{i}{2} \int_0^T \!dt\int dV \psi_{\pmb{k}}^* \Big(\theta(z-Z_{\text{I}}) \left( \Psi \partial_{\rho} \delta(R_{\text{I}}-\rho) + 2 (\partial_{\rho} \Psi)\delta(R_{\text{I}}-\rho) +\rho^{-1} \Psi  \delta(R_{\text{I}}-\rho) \right)\nonumber\\
&& \qquad - \theta(R_{\text{I}}-\rho) \left( \Psi \partial_{z} \delta(z-Z_{\text{I}}) +2 (\partial_{z} \Psi)\delta(z-Z_{\text{I}}) + 2iA(t) \Psi  \delta(z-Z_{\text{I}}) \right) \Big)\nonumber\\
&=& \frac{i}{2} \int_0^T \!dt\int_{0}^{2\pi}\!d\varphi \bigg( \int_{Z_{\text{I}}}^{\infty} dz \left( -\partial_{\rho}(\rho\psi_{\pmb{k}}^*\Psi) + 2\rho \psi_{\pmb{k}}^*(\partial_{\rho} \Psi) +\psi_{\pmb{k}}^* \Psi \right)|_{\rho=R_{\text{I}}}\nonumber\\
&& \qquad - \int_{0}^{R_{\text{I}}} \rho d\rho \left( -\partial_{z}(\psi_{\pmb{k}}^*\Psi) +2 \psi_{\pmb{k}}^* (\partial_{z} \Psi) + 2iA(t)\psi_{\pmb{k}}^*  \Psi  \right)|_{z=Z_{\text{I}}} \bigg)\; .
\end{eqnarray}
The second term is, analogously,
\begin{multline}
  \label{eq:summand_2}
  s_2(k_{\rho},k_z)
= \frac{i}{2} \int_0^T \!d t\int_{0}^{2\pi}\!d\varphi \bigg( \int_{-\infty}^{-Z_{\text{I}}} d z \left( -\partial_{\rho}(\rho\psi_{\pmb{k}}^*\Psi) + 2\rho \psi_{\pmb{k}}^*(\partial_{\rho} \Psi) +\psi_{\pmb{k}}^* \Psi \right)|_{\rho=R_{\text{I}}}\\
+ \int_{0}^{R_{\text{I}}} \rho \, d \rho \left( -\partial_{z}(\psi_{\pmb{k}}^*\Psi) +2 \psi_{\pmb{k}}^* (\partial_{z} \Psi) + 2iA(t)\psi_{\pmb{k}}^*  \Psi  \right)|_{z=-Z_{\text{I}}} \bigg),
\end{multline}
and 
\begin{eqnarray}
  \label{eq:summand_3}
  s_3(k_{\rho},k_z) &=& \int_0^T \!d t\int\!dV \partial_t\Big(\psi_{\pmb{k}}^* \theta(\rho-R_{\text{I}}) \Psi \Big) \nonumber \\
  &=&\frac{-i}{2} \int_0^T \!d t\int_{0}^{2\pi}\!d\varphi \int_{-\infty}^{\infty} d z  \left( -\partial_{\rho} (\rho \Psi \psi_{\pmb{k}}^*) + 2\rho \psi_{\pmb{k}}^* (\partial_{\rho} \Psi) +\psi_{\pmb{k}}^*  \Psi  \right)|_{\rho=R} .
\end{eqnarray}
Inserting the Volkov wavefunction
\begin{equation}
  \label{eq:volkov}
  \psi_{\pmb{k}}(\rho,z,t)=(2\pi)^{-3/2} \eulere^{-\imagi\alpha(t)+\imagi k_z z +\imagi k_{\rho} \rho\cos\varphi}, \qquad \alpha(t) = \frac{1}{2}\int_0^t d t'\, [k^2+2k_zA(t')]
\end{equation}
and collecting all integrals over $z$ in $s'_3$ such that $s'_1+s'_2+s'_3=s_1+s_2+s_3$ yields
\begin{equation}
  \label{eq:s_1_volkov_wo_phi}
  s'_1(k_{\rho},k_z) = - \frac{i}{2} (2\pi)^{-1/2} \eulere^{-\imagi k_zZ_{\text{I}}}\int_0^T \!d t \eulere^{\imagi\alpha(t)}  \int_0^{R_{\text{I}}} \rho \, d \rho  J_0(k_{\rho} \rho) \left(\imagi k_z\Psi  + \partial_z \Psi + 2iA \Psi \right)|_{z=Z_{\text{I}}}\; ,
\end{equation}
\begin{equation}
  \label{eq:s_2_volkov_wo_phi}
  s'_2(k_{\rho},k_z) = \frac{i}{2} (2\pi)^{-1/2} \eulere^{\imagi k_zZ_{\text{I}}}\int_0^T \!d t \eulere^{\imagi\alpha(t)}  \int_0^{R_{\text{I}}} \rho \, d \rho  J_0(k_{\rho} \rho) \left(\imagi k_z\Psi  + \partial_z \Psi + 2iA \Psi \right)|_{z=-Z_{\text{I}}},
\end{equation}
and
\begin{equation}
  \label{eq:side_integral_wo_phi}
  s'_3(k_{\rho},k_z) = \frac{-i}{2} (2\pi)^{-1/2} R_{\text{I}} \int_0^T \!d t \eulere^{\imagi\alpha(t)}  \int_{-Z_{\text{I}}}^{Z_{\text{I}}}  d z \eulere^{-\imagi k_z z} \left( \Psi  k_{\rho}  J_1(k_{\rho} R_{\text{I}}) + J_0(k_{\rho}R_{\text{I}}) \partial_{\rho} \Psi  \right)|_{\rho=R_{\text{I}}} %\; .
\end{equation}
where $J_0$ and $J_1$ are Bessel functions of the first kind.
Denoting the Fourier transform
\begin{equation}
  \label{eq:fourier_trafo}
  \Psi(\rho,k_z,t) = \int_{-Z_{\text{I}}}^{Z_{\text{I}}}  d z \eulere^{-\imagi k_z z}  \Psi(\rho,z,t)
\end{equation}
and the Hankel transform
\begin{equation}
  \label{eq:hankel_trafo}
  \Psi(k_{\rho},z,t) = \int_0^{R_{\text{I}}}  \, d \rho \, \rho J_0(k_{\rho} \rho) \Psi(\rho,z,t)
\end{equation}
the approximation of the probability amplitude reads
\begin{multline}
  \label{eq:result_cylinder_tsurff}
s_1+s_2+s_3= \frac{-i}{2\sqrt{2\pi}}  \int_0^T \!d t \eulere^{\imagi\alpha(t)}\big( \eulere^{-\imagi k_zZ_{\text{I}}} \left(\imagi k_z  + \partial_z  + 2iA \right) \Psi(k_{\rho},z,t)|_{z=Z_{\text{I}}}
- \eulere^{\imagi k_zZ_{\text{I}}} \left(\imagi k_z  + \partial_z  + 2iA \right) \Psi(k_{\rho},z,t)|_{z=-Z_{\text{I}}}\\ 
+ R_{\text{I}}  \left( k_{\rho}  J_1(k_{\rho} R_{\text{I}}) + J_0(k_{\rho}R_{\text{I}}) \partial_{\rho} \right)\Psi(\rho,k_z,t)|_{\rho=R_{\text{I}}} \big)\; .
\end{multline}
The GSL \cite{gsl} was used for the Hankel transform and the Bessel functions.

In order to suppress spurious effects introduced by the finite time $T$ in the t-SURFF time integrals a Hanning window
\begin{equation}
  \label{eq:hanning_window}
H(t)=
  \begin{cases}
   1 & \text{if } t < T/2 \\
   [1-\cos(2\pi t/T)]/2       & \text{if } t \geq T/2
  \end{cases}  
\end{equation}
was multiplied to the integrands (\ref{eq:DI_integral_final}), (\ref{eq:ion_amplitude_oneD}), and (\ref{eq:result_cylinder_tsurff}).

\begin{table}%[H] add [H] placement to break table across pages
\caption{\label{tab:parameters}Parameters for the numerical simulations. All values in atomic units.}
\begin{ruledtabular}
\begin{tabular}{lll}
system & spatial grid & t-SURFF\\
\colrule
2D & $\Delta x=0.1$, $\Delta R=0.01$, & $X_{\text{DI}}=110$\\
 & $x\in[-300,300)$, $R\in(0,20]$ & \\
fixed $R$ (1D)& $\Delta x=0.1$, $x\in[-300,300)$ & $X_{\text{I}}=100$\\
fixed $R$ (3D)& $\Delta z=0.1$, $z\in[-200,200)$ & $Z_{\text{I}}=150$\\
& $\Delta \rho=0.05$, $\rho\in(0,100]$ & $R_{\text{I}}=50$\\
\colrule
\multicolumn{3}{c}{for all simulations: $\Delta t=0.0125$, $T=3T_{\text{p}}$}\\
\end{tabular}
\end{ruledtabular}
\end{table}

\end{widetext}

\end{document}